    \documentclass[twocolumn]{aastex631}

    \usepackage{graphicx}
    \usepackage{amsmath}
    \usepackage{natbib}
    \usepackage{booktabs}
    \newcommand{\dgr}{^{\circ}}
    \newcommand{\omdot}{\dot{\Omega}}

    \newcommand*{\rd}[2]{\frac{\mathrm{d}#1}{\mathrm{d}#2}}
    \newcommand*{\pd}[2]{\frac{\partial#1}{\partial#2}}
    \newcommand*{\rdil}[2]{\mathrm{d}#1 / \mathrm{d}#2}
    
    \newcommand*{\rtd}[2]{\frac{\mathrm{d}^2#1}{\mathrm{d}#2^2}}

    \renewcommand*{\bm}[1]{\boldsymbol{\mathbf{#1}}}
    \newcommand*{\uv}[1]{\hat{\bm{#1}}}

    \newcommand*{\abs}[1]{\left|#1\right|}
    
    \newcommand*{\p}[1]{\left(#1\right)}
    \newcommand*{\s}[1]{\left[#1\right]}
    \newcommand*{\z}[1]{\left\{#1\right\}}

    \renewcommand{\b}[1]{\boldsymbol{#1}} 
    \newcommand{\ihat}{\hat{\textbf{\i}}} 
    \newcommand{\jhat}{\hat{\textbf{\j}}}
    \newcommand{\khat}{\bm{\hat k}}
    \newcommand{\xhat}{\bm{\hat x}}
    \newcommand{\yhat}{\bm{\hat y}}
    \newcommand{\zhat}{\bm{\hat z}}
    \renewcommand{\b}[1]{\boldsymbol{#1}}
    
    \newcommand{\tnpar}{\tau_{\textsc{npar}}}
    \newcommand{\Wnpar}{\Omega_{\textsc{npar}}}

    \received{XXXX}
    \revised{YYYY}
    \accepted{ZZZZ}

    \submitjournal{ApJ}

\begin{document}

\title{Surprising Spin-orbit Resonances of Rocky Planets}
\shorttitle{Spin-orbit Resonances of Super-Earths}

\correspondingauthor{Yubo Su}
\email{yubosu@princeton.edu}
\author{Henry D. A. Yuan}
\affiliation{Department of Astrophysical Sciences, Princeton University, Princeton, NJ 08544, USA}
\author[0000-0001-8283-3425]{Yubo Su}
\affiliation{Department of Astrophysical Sciences, Princeton University, Princeton, NJ 08544, USA}
\author[0000-0002-6710-7748]{Jeremy Goodman}
\affiliation{Department of Astrophysical Sciences, Princeton University, Princeton, NJ 08544, USA}


\begin{abstract}


Recent works suggest that, in multiplanetary systems, a close-in exoplanet can
sometimes avoid becoming tidally locked to its host star if it is captured into
a secular spin-orbit resonance with a companion planet. In such a resonance, the
planet remains at a sub-synchronous spin rate and an appreciable obliquity (the
planet's spin-orbit misalignment angle). However, many of these works have only
considered planets with fluid-like rheologies. Recent observations suggest that
planets up to a few Earth masses may be rocky and thus may have an appreciable
rigidity. In this work, we study the spin-orbit dynamics of such rigid planets
using a linear dissipative tidal model and not enforcing principal axis rotation
about the body's shortest principal axis. We identify a new class of spin-orbit
resonances when the planet spins at twice its orbital frequency. These resonances exist at nonzero obliquity and spontaneously excite
non-principal-axis rotation upon resonance capture. While these resonances
eventually disappear as tidal dissipation damps the obliquity to zero (and the body returns to principal-axis rotation), they
still modify the spin evolutionary history of the planet. Such resonances may enhance the prevalence of secular spin-orbit resonances in exoplanetary systems.

\end{abstract}

\section{Introduction}\label{s:intro}

The obliquity of a planet, the angle between its rotational
and orbital axes, plays a significant role in the planet's evolution. First, the
obliquities of exoplanets are thought to affect their potential habitability
\citep[e.g.][]{williams1997, spiegel2009, heller2011, armstrong2014}. For
instance, Earth's $23\dgr$ obliquity is responsible for the temperate seasons we
experience at lower latitudes, while a planet with an obliquity greater than
$54\dgr$ instead experiences greater time-averaged insolation at the poles than
at the equator \citep{ferreira2014, lobo2020}. In addition, planets that retain
large obliquities in the presence of tidal dissipation rotate subsynchronously
\citep{levrard2007, fabrycky2007}, avoiding the so-called ``tidally locked''
state (spin-orbit synchronization) that results in climatic effects that might
be hostile to life \citep[e.g.][]{kite-etal2011, penn-vallis2017}. Such concerns
are of particular interest for planets orbiting M dwarfs in their habitable
zones, which are generally assumed to be tidally locked \citep[e.g.][]{yang2014}
but many of which may retain large obliquities \citep{2022valenteobl,
guerrero2023}.

Second, the obliquities of planets also affect their dynamical histories, as
tidal dissipation in oblique planets can be enhanced significantly compared to
that in aligned planets, resulting in different orbital evolution
\citep[e.g.][]{millholland_wasp12b, millholland2019obliquity,
millholland2020formation, su-lai2022-res}. While the direct measurement of
exoplanetary obliquities is difficult, constraints on the spin-orbit
misalignments of directly imaged distant Jupiter- and super-Jupiter-mass
companions have recently been obtained \citep{bryan2020obliquity,
bryan2021obliquity}, and the determination of the spin properties of close-in
exoplanets might soon be possible \citep[e.g.][]{seager2002constraining,
carter2010, snellen2014fast, ohno2019, adams2019}, such as the recent claim to
have inferred tidal locking for an ultra-short-period (0.46 day) rocky
super-Earth \citep{kreidberg2019, lyu2023}.

The study of the origins and evolutions of planetary spins began as an effort to
explain the diverse spin states of the planets and satellites within the solar
system. Famously, the $98\dgr$ obliquity of Uranus is typically attributed to a
giant impact \citep{benz1989, korycansky1990, slattery1992, ida2020}, a
mechanism that has also been applied to explain Neptune's $28\dgr$ obliquity
\citep[e.g.][]{reinhardt2020}. However, another class of mechanisms that has
been invoked to explain the spin states of many other solar system bodies is
\emph{spin-orbit resonance}. One kind of spin-orbit resonance (``non-secular'')
is exemplified by Mercury's well-known 3:2 ratio of its spin and orbital
frequencies, which is a result of its eccentric orbit and permanent azimuthal
asymmetry \citep{colombo1966}; spin states where the spin frequency is a
half-integer multiple of the orbital frequency are generally possible for
triaxial bodies, which have three distinct moments of inertia
\citep{goldreich-peale1966, correia-laskar2004}. A second kind of spin-orbit
resonance (``secular'') is a commensurability of secular spin and orbital
precession frequencies.
Secular resonances have been invoked to explain Saturn's $26.7\dgr$
obliquity \citep{ward-hamilton2004-Ianalytic, hamilton-ward2004-IInumeric,
saillenfest-etal2021-migration}, Jupiter's $3.1\dgr$ obliquity
\citep{ward-canup2006, saillenfest-etal2020, Dbouk2023}, and even those of
Uranus and Neptune again \citep{rogoszinski2020, rogoszinski2021,
saillenfest-etal2022, lu2022_p9}. The overlap of multiple such resonances is
thought to give rise to the chaotic obliquity evolution of Mars
\citep{touma1993chaotic, laskar1993chaotic}. The theory for secular spin-orbit
resonances dates back to a generalization of the results of \citet{cassini1693}
to more general systems \citep{colombo1966, peale1969, peale1974,
henrard-murigande1987}.

When applying the results of these studies to the dynamics of extrasolar
planetary systems, a new complication arises: many of the known exoplanets are
found in much shorter-period orbits than those of the solar system planets \citep[e.g.\ the
``Kepler multis'',][]{borucky2011, fabrycky2014}, where they can experience
tidal damping of their obliquities (and tidal spin-orbit synchronization). By
contrast, in the solar system only Mercury and Venus have tidal dissipation
timescales shorter than the age of the solar system \citep{lissauer2012}. Early
works on the combined effect of spin-orbit resonances and tides include the
demonstrations that both non-secular and secular spin-orbit resonances are
tidally stable \citep{colombo1966, ward1975}. The interaction of secular
spin-orbit resonances and tidal dissipation in exoplanetary systems was expanded
by \citet{su-lai2022-res}. They studied super Earths (SEs), which are thought to
be formed by a stage of late giant impacts and hence to have a broad initial
obliquity distribution \citep{dones1993, inamdar2015formation}, and which are
thought to be accompanied by outer cold Jupiters (CJs) $\sim 30\%$ of the time
\citep{bryan2018constraints, zhu2018super}. In planetary systems containing SEs
with CJ companions, \citet{su-lai2022-res} found that tidal damping of the large
primordial SE obliquities causes a substantial fraction ($\sim 30\%$) of them to
be trapped in a high-obliquity secular spin-orbit resonance called \emph{Cassini
State 2} (after the nomenclature of \citealp{peale1969}). Later work found that
large SE obliquities can occur in multiplanetary systems as well, as tidal
dissipation tames the chaotic obliquity evolution, and is even somewhat enhanced
due to higher-order secular spin-orbit resonances \citep{saillenfest-etal2019,
su-lai2022-obl}.

However, the results of \citet{su-lai2022-res} and many contemporaneous works
\citep[e.g.][]{millholland2019obliquity, lu2023rebound} assume that their
planets deform hydrostatically, where the planet's figure and spin state
are simply related to each other.
While this is likely a good approximation for planets with massive
gaseous envelopes, recent works have found that SEs are likely rocky up to a few
Earth masses \citep[e.g.][]{carter2012, howard2013, fulton2017, owen2019,
otegi-etal2020, luque2022}, though other works have suggested that many such
planets may be ``water worlds'' with a substantial mass fraction of solid or
liquid water \citep{zeng2019, rogers2023}; see \citet{bean2021} for good
reviews. Since these SEs will have some inherent rigidity that
contributes to their figure, it is important and timely to characterize the spin
evolution of planets with rigid shapes.


In this work, we mostly adopt and expand upon the model of \citet{gladman1996}
due to its simplicity and qualitative accuracy.
In their work, they assume that the rocky body retains a rigid
shape, remains in principal-axis rotation (rotation about an eigenvector of the
planet's moment of inertia tensor, typically referring to rotation about the
shortest axis), and experiences tidal dissipation via the standard constant time
lag (CTL) model \citep{alexander1973weak, mignard1979, hut1981tidal}. In our
work, we expand on the second of these three assumptions and show that new
dynamics arise the planet is allowed to deviate from principal-axis rotation,
even when the deviation is damped in a self-consistent way.
Regarding the third of these assumptions, note that tidal dissipation in
rocky bodies likely differs significantly from the CTL model, and it may be better modeled by the Maxwell and Andrade
rheologies among others \citep{dobrovolskis1980, efroimsky2009,
efroimsky2012_tides, ferraz2013, delisle2017, correia2022formalism}.
However,
since the detailed tidal dissipation in planets can depend on many factors (such
as oceans on the Earth), we adopt the widely-used CTL model for this work (see
additional discussion in Section~\ref{sec:disc}). In Section~\ref{sec:var&eq},
we introduce the equations governing the planet's spin evolution. In
Section~\ref{sec:num}, we describe the results of numerical integrations of
these equations. In Section~\ref{sec:ana}, we present a Hamiltonian analysis of
the resonances found in Section~\ref{sec:num}. We summarize our results and
discuss in Section~\ref{sec:disc}.

\section{Variables \& Equations}\label{sec:var&eq}

Our problem consists of a star of mass $M_\star$ that hosts a planet on
a circular orbit with semi-major axis $a$, mass $m$, radius $R$, spin vector
$\bm{\Omega}$, and orbit normal $\hat{\bm{\ell}}$. We denote the instantaneous
radial vector from the planet to the star by $\bm{r}$. While $\bm{r}$ is
often defined as the vector from the star to the planet, we define it in this
way because we are interested in the planet's spin dynamics, so we center our
coordinates on the planet. We define two separate coordinate systems: an
inertial system with orthonormal basis $\{\xhat,\yhat,\zhat\}$, $\zhat$ along
the orbit normal; and a body-fixed set of coordinates centered on the planet,
with orthonormal basis $\{\ihat,\jhat,\khat\}$. The latter vectors lie along the
planet's three principal axes, with respective moments of inertia
$A\leq B\leq C$. We refer to these two coordinate systems hereafter as $\{xyz\}$
and $\{ijk\}$ respectively.

In this paper, we are interested in the spin dynamics of rigid bodies, so we
assume that the three principal moments of inertia of the body are constant. We
introduce the normalized moment of inertia $k$ (not to be confused with either the Love number $k_2$ or the body unit vector $\khat$), the triaxiality $\eta_{\rm
tri}$, and the oblateness $\eta_{\rm obl}$ as
\begin{align}
    \label{eq:k}
    k &\equiv \frac{A}{mR^2},
    \\
    \label{eq:eta_tri}
    \eta_{\rm tri} &\equiv \frac{B-A}{C},
    \\
    \label{eq:eta_obl}
    \eta_{\rm obl} &\equiv \frac{C-A}{C}.
\end{align}
For a sphere of uniform density, $k = 2/5$ and $\eta_{\rm tri} = \eta_{\rm obl}
= 0$. Note that if $\eta_{\rm tri} \ll \eta_{\rm obl}$, then $\eta_{\rm
obl}\approx J_2/k$, $J_2$ being the dimensionless gravitational quadrupole
moment. Note that the hydrostatic contribution to $\eta_{\rm
obl}$, given by
\begin{equation}
    \eta_{\rm obl, hs} = \frac{k_2}{3k}\frac{\Omega^2}{GM/R^3},
\end{equation}
will be less than our adopted $\eta_{\rm obl}$ for spin periods $\gtrsim
10\;\mathrm{day}$ for rocky planets.

We next define a few quantities that will facilitate discussion throughout
this paper. The obliquity, $\theta$, is the angle between the spin vector and
the orbit normal,
\begin{equation}\label{eq:theta}
    \cos{\theta} \equiv \hat{\bm{\Omega}}\cdot\hat{\bm{\ell}},
\end{equation}
where $\hat{\bm{\Omega}} = \bm{\Omega}/\Omega$. Next, $\beta$ is defined as the angle
between $\hat{\bm{\Omega}}$ and $\pm\khat$, defined by
\begin{equation} \label{eq:beta}
    \cos{\beta} \equiv \pm\hat{\bm{\Omega}} \cdot\khat = \pm s_k,
\end{equation}
where the sign is chosen to ensure $0\leq\beta\leq\frac{\pi}{2}$. The case
$\beta=0$ corresponds to principal-axis rotation about the shortest axis
(largest moment). We do not enforce principal-axis rotation in our numerical
implementation presented in Section~\ref{sec:num}, so this angle can generally
be nonzero.

\subsection{Spin Dynamics}\label{ss:dyn}

The evolution of $\bm{\Omega}$ depends on the torque exerted on the planet by
the star. Following \citep{gladman1996}, we first consider two dominant torques.
The first is the torque exerted on the planet's asymmetrical shape, hereafter
referred to as the rigid-body torque. This torque is given by MacCullagh's
formula \citep{gladman1996}:
\begin{align}
    \bm{\Gamma}_{\rm RB} ={}& \frac{3GM_\star}{r^5}\Bigr[
        (C-B) (\bm{r}\cdot\jhat)
            (\bm{r}\cdot\khat) \ihat\nonumber\\
        &+ (A-C) (\bm{r}\cdot\khat)
            (\bm{r}\cdot\ihat)\jhat\nonumber\\
        &+ (B-A)(\bm{r}\cdot\ihat)
            (\bm{r}\cdot\jhat)\khat
    \Bigr].\label{eq:rbtriax}
\end{align}

The second torque component is exerted on the tidally deformed shape of the
planet, referred to hereafter as the tidal dissipative torque. We use the
constant time lag tidal model \citep{alexander1973weak, mignard1979, hut1981tidal}, which
assumes that the tidal bulge lags the location of the perturber by a constant
time offset $\tau$. $\tau$ does not evolve in time and is independent of the
forcing frequency $(\Omega - n)$. Under this model, the tidal dissipative torque
is given by \citep{gladman1996}:
\begin{equation} \label{eq:tidaltorque}
    \bm{\Gamma}_{\rm tide} = \frac{3k_2GM_\star^2R^5}{r^6}
        \frac{(\bm{r}\cdot\bm{\rho})(\bm{\rho}\times\bm{r})}{r^2\rho^2}.
\end{equation}
$k_2$ and $R$ are the second degree Love number and mean radius of the planet
respectively, and $\bm{\rho}$ is the vector from the planet to the star's
retarded position, time-lagged by $\tau$ behind $\bm{r}$ in the planet's
body-fixed coordinates. $\bm{\rho}$ is given to leading order in $\tau$ by
\citep{gladman1996}\footnote{\citet{gladman1996} assume principal axis rotation,
such that $\bm{\Omega}$ is always aligned with $\khat$. We do not make this
assumption, allowing $\hat{\bm{\Omega}}$ and $\khat$ to evolve independently. We
therefore change Eq.~\eqref{eq:rho} slightly to account for this difference,
replacing $\Omega\khat$ with $\bm{\Omega} = \Omega\hat{\bm{\Omega}}$.}
\begin{equation}\label{eq:rho}
    \bm{\rho} \approx \bm{r}-\bm{\dot r}\tau+(\bm{\Omega}\times\bm{r})\tau,
\end{equation}
where $\tau$ is chosen by association with an effective tidal quality factor $Q$
\citep{goldreich-soter1966, lu2023rebound}
\begin{equation}\label{eq:tidal_Q}
    \frac{1}{Q} = \tan{\left(2n\tau\right)}.
\end{equation}
Since we've assumed the planet's orbit is circular, we have that
$\dot{\bm{r}}=n\hat{\bm{\ell}}\times\bm{r}$, $n$ being the mean motion. Note
that this corresponds to a characteristic tidal spin evolutionary rate of
\begin{equation}
    -\rd{\ln \theta}{t} \sim \frac{1}{\tau_{\rm tide}}
        =
            \frac{3}{2k}
            \frac{k_2}{Q}
            \frac{M_\star}{m}
            \p{\frac{R}{a}}^3
            n\label{eq:tau_tide}.
\end{equation}

In addition to these two torques, we also introduce a third effect. While
we do not enforce principal axis rotation (PAR) in this study, bodies in the
solar system are observed to be in PAR, presumably due to rapid damping of non-PAR motion
\citep{burns1973, peale1973}. We adapt Eq.~(56) of \citet{peale1973} (see
Appendix~\ref{app:goodman_damping} for an approachable derivation of this
expression) into the following non-PAR damping torque:
\begin{align}
    \bm{\Gamma}_{\rm NPAR}
        &= \Gamma_{\rm NPAR}
            \frac{\s{\uv{\Omega} \times \p{
            \khat \times
            \uv{\Omega}}}}{\sin\beta}
            \label{eq:npar_torque}
            ,\\
    \Gamma_{\rm NPAR} &\equiv
        \frac{1}{3}\frac{k_2}{Q}\frac{\Omega^4R^5}{G}\sin\beta\cos^2\beta.
\end{align}
This torque acts to drive $\bm{\Omega}$ towards $\bm{k}$ at the characteristic
rate
\begin{align}
    -\rd{\ln \beta}{t} \sim \frac{1}{\tau_{\rm NPAR}} &=
        \frac{1}{3k}
        \frac{k_2}{Q}
        \frac{\Omega_k^2}{Gm/R^3}\Omega_k
        \label{eq:tau_npar}
        \\
         &= \frac{2}{9}\frac{1}{\tau_{\rm tide}}\p{\frac{\Omega_k}{n}}^3\nonumber.
\end{align}
In the case of the Earth, where $\Omega_k\approx 366n$, Eqs.~(\ref{eq:tau_tide}) and~\eqref{eq:tau_npar} give
that $\tau_{\rm tide} \sim \;\mathrm{Gyr}$ and $\tau_{\rm NPAR} \sim
20\;\mathrm{yr}$, consistent with the values found in
\citet{peale1973}\footnote{Interestingly, the Earth is observed to exhibit free
precession even today, this motion being termed the \emph{Chandler wobble}. Due
to the expected rapid damping of this free precession, it must be continually
excited, and the leading theory is a combination of atmospheric and oceanic
processes \citep{gross2000}.}. While this is indeed very short,
Eq.~\eqref{eq:tau_npar} shows that NPAR damping is actually a factor of $9/2$
\emph{slower} than tidal realignment for planets with $\Omega \sim n$, i.e.\
near spin-orbit synchronization \citep[as has also been pointed out in
e.g.][]{gladman1996}.

Given the net torque vector on a body in the body-fixed $\{ijk\}$ coordinates
$\bm{\Gamma} = \bm{\Gamma}_{\rm RB} + \bm{\Gamma}_{\rm tide}$, the evolution of
the spin vector $\bm{\Omega}$ is governed by Euler's rotation equations. Using
subscripts $i$, $j$, $k$ to denote the components of a vector along the three
body axes, these equations are \citep{goldstein2002}
\begin{align}
    \label{eq:euler1}
    A\omdot_i + (C-B)\Omega_j\Omega_k &= \Gamma_i,
    \\
    \label{eq:euler2}
    B\omdot_j + (A-C)\Omega_k\Omega_i &= \Gamma_j,
    \\
    \label{eq:euler3}
    C\omdot_k + (B-A)\Omega_i\Omega_j &= \Gamma_k.
\end{align}
Then, as the planet rotates, the $\{\ihat \jhat \khat\}$ basis vectors evolve with respect
to the fixed $\{\uv{x}\uv{y}\uv{z}\}$ basis. These unit vectors rotate about the axis
$\hat{\bm{\Omega}}$ at spin rate $\Omega$
\begin{equation} \label{eq:dijk-dt}
    \frac{d\hat{\bm{e}}}{dt} = \bm{\Omega} \times \hat{\bm{e}},
\end{equation}
where $\hat{\bm{e}} \in\{\ihat,\jhat,\khat\}$.

\section{Numerical Results}\label{sec:num}

We use the N-body code REBOUND \citep{rebound} to simultaneously evolve the
planet's orbit and spin. REBOUND's N-body integration handles the orbital
component, and we integrate Eqs.~(\ref{eq:euler1}--\ref{eq:dijk-dt}) in parallel
with REBOUND's arbitrary ODE solver, which uses an adaptive
Gragg-Bulirsch-Stoer integrator
\citep{hairer-norsett-wanner1993, rebound}.\footnote{A high-order integrator is
required to capture the correct dynamics near $\theta=90\dgr$; we used a
lower-order integrator with a fixed timestep in an earlier iteration of this
work, which led to spurious oscillations about $\theta=90\dgr$.}
The order and timestep of the integrator are automatically adjusted by REBOUND
to meet a specified absolute and relative tolerance of $10^{-8}$.

\begin{table}[h!]
\centering
\begin{tabular}{ |p{1.5cm}|p{1.8cm}||p{1.5cm}|p{1.8cm}| }
 \midrule
 \multicolumn{4}{|c|}{Simulation Parameter Values} \\
 \midrule
 Parameter & Value & Parameter & Value\\
 \midrule
     $M_\star$ & $1.2$ $M_{\odot}$    & $k_2$             & $0.3$\\
     $m$       & $1.57$ $m_{\oplus}$  & $k$               & $0.331$\\
     $a$       & $0.124$ AU           & $\eta_{\rm tri}$  & $10^{-6}$ or $0$\\
     $R$       & $1.176$ $R_{\oplus}$ & $\eta_{\rm obl}$  & $10^{-5}$ \\
     $Q$       & $300$                & $\beta$ (Initial) & $0\dgr$\\
 \midrule
\end{tabular}
\caption{Parameter values for numerical simulations, based on the parameters of
the super-Earth Kepler-1501b and Earth itself. All of these parameters are fixed
throughout a given simulation except $\beta$ [Eq.~\ref{eq:beta}], which is
initialized as shown.}\label{tab:sim-params}
\end{table}

\subsection{Parameter Choices}

For the parameters of the system, we adopt those of
Kepler-1501b as an archetypal rocky SE, with parameters listed in
Table~\ref{tab:sim-params}\citep{kepler1501mass, nasa_exoarchive}, and we
estimate the mass using the mass-radius relation of \citep{otegi-etal2020}. For
parameters unconstrained by observation, we choose their values roughly based on
the estimated values for Earth: $k_2=0.3$, $Q=300$, and $k=0.331$,
\citep{yoder1995book, lainey2016}. We also initialize the planet to be in
principal-axis rotation about its shortest axis (i.e. $\hat{\Omega}=\hat{k}$) such
that $\beta=0\dgr$. For these parameters, the characteristic spin evolution
timescale is obtained by evaluating Eq.~\eqref{eq:tau_tide}
\begin{align}
    \frac{\tau_{\rm tide}}{P}
    \approx{}& 2.1 \times 10^6
        \p{\frac{k_2/Q}{10^{-3}}}^{-1}
        \frac{\Omega}{n}
        \p{\frac{M_\star}{1.2 M_{\odot}}}^{-1}\nonumber\\
        &\times \p{\frac{m}{1.57 M_{\oplus}}}
        \p{\frac{R}{1.176 R_{\oplus}}}^{-3}
        \p{\frac{a}{0.124\;\mathrm{AU}}}^{3},
\end{align}
where $P$ is the orbital period of the planet. As such, we choose the length of
our numerical integrations to be $10^7 P$, such that the planet has reached a
tidal equilibrium state. Note also that for these parameters, the orbital decay
rate of the planet (both due to tides raised on the planet and on the star) is
$\gg \;\mathrm{Gyr}$ \citep{2012Lai}, so we can safely neglect its orbital
decay.

As for the shape of the planet, its maximum asphericity scales
with the dimensionless \emph{effective rigidity} \citep{murray_dermott_1999,
zanazzi-lai2017}
\begin{equation}
    \tilde{\mu} = \frac{19 \mu}{2\rho g R} \propto \frac{R^4}{M^2},
        \label{eq:rigidity}
\end{equation}
where $g = GM/R^2$ is the surface acceleration and $\mu$ is the shear modulus of
the planet. While $\mu$ for the Earth is measured to be
$10^{12}\;\mathrm{dyn/cm^2}$ \citep{turcotte2002}, the exact relation between
$\tilde{\mu}$ and the maximum asphericity is somewhat nontrivial to estimate
\citep[e.g.][]{zanazzi-lai2017}. However, we can scale the deformation to
measurements of Venus's oblateness, assuming that the bulk modulus $\mu$ is
similar for all rocky planets: since Venus's $J_2$ moment exceeds its
hydrostatic value by a factor of $25$ \citep{yoder1995, dumoulin2017venus}, its
measured $\eta_{\rm obl} \approx 1.3 \times  10^{-5}$ and $\eta_{\rm tri} \sim 6
\times 10^{-6}$ \citep{yoder1995book} are likely supported by its inherent
rigidity. Noting that the ratio $R^4/M^2$ is similar for Venus and Kepler-1501b,
we adopt $\eta_{\rm obl} \sim 10^{-5}$ and $\eta_{\rm tri} \sim 10^{-6}$.
Scaling to Earth's excess oblateness (above its hydrostatic value) and its triaxiality, which are both $\sim 10^{-5}$ \citep{yoder1995book}, gives similar results.


\begin{figure}
    \centering
    \includegraphics[width=0.8\columnwidth]{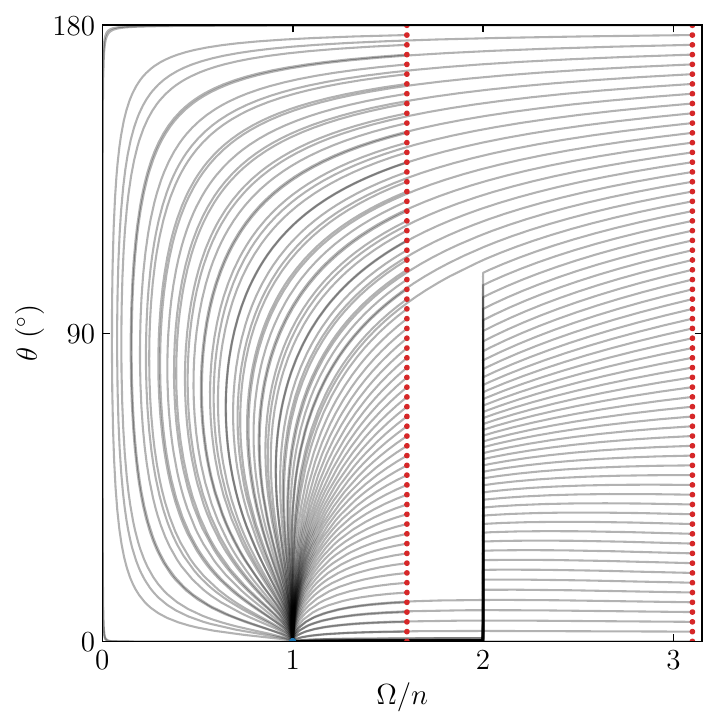}
    \caption{The spin evolution of an oblate Kepler-1501b-like super Earth in the
    coordinate space consisting of $\Omega / n$ (the ratio of the spin rate to
    the orbital frequency) and $\theta$ (the obliquity, the angle between the
    planet's spin axis and its orbit normal) using the system parameters given
    in Tab.~\ref{tab:sim-params} (with $\eta_{\rm tri} = 0$). $128$ integrations
    are displayed, and each is integrated for $3 \times 10^7$ orbital periods. The
    initial conditions are marked in red, while the final coordinates are marked
    in blue. While all trials ultimately converge to $\theta = 0$ and $\Omega/n
    = 1$, many evolutionary tracks are affected by the 2:1 spin-orbit
    resonance.}\label{fig:obl-grid}
\end{figure}
\begin{figure}
    \centering
    \includegraphics[width=0.8\columnwidth]{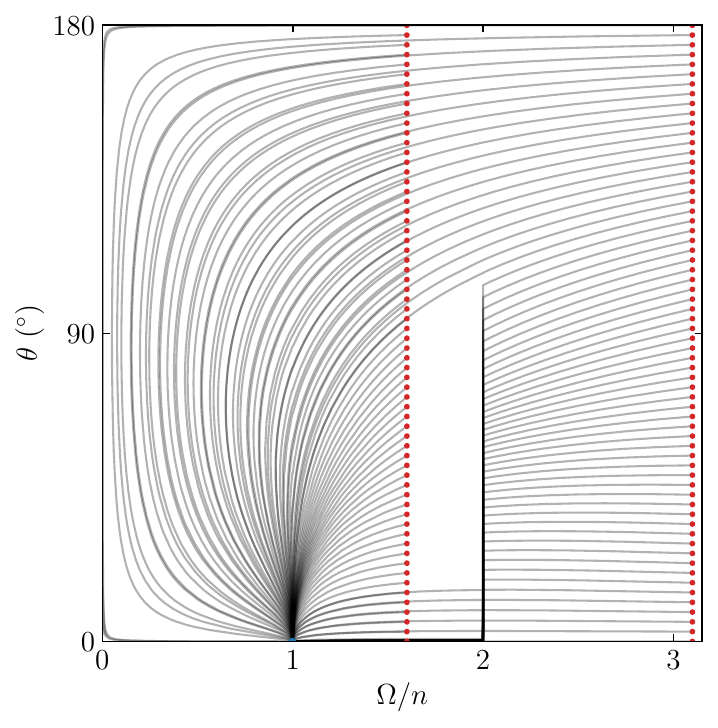}
    \caption{Same as Fig.~\ref{fig:obl-grid} but for $\eta_{\rm tri} = 10^{-5}$;
    while the detailed capture into the 2:1 resonances changes slightly, the
    qualitative evolutionary features are unaffected.}\label{fig:triax-grid}
\end{figure}
\begin{figure}
    \centering
    \includegraphics[width=\columnwidth]{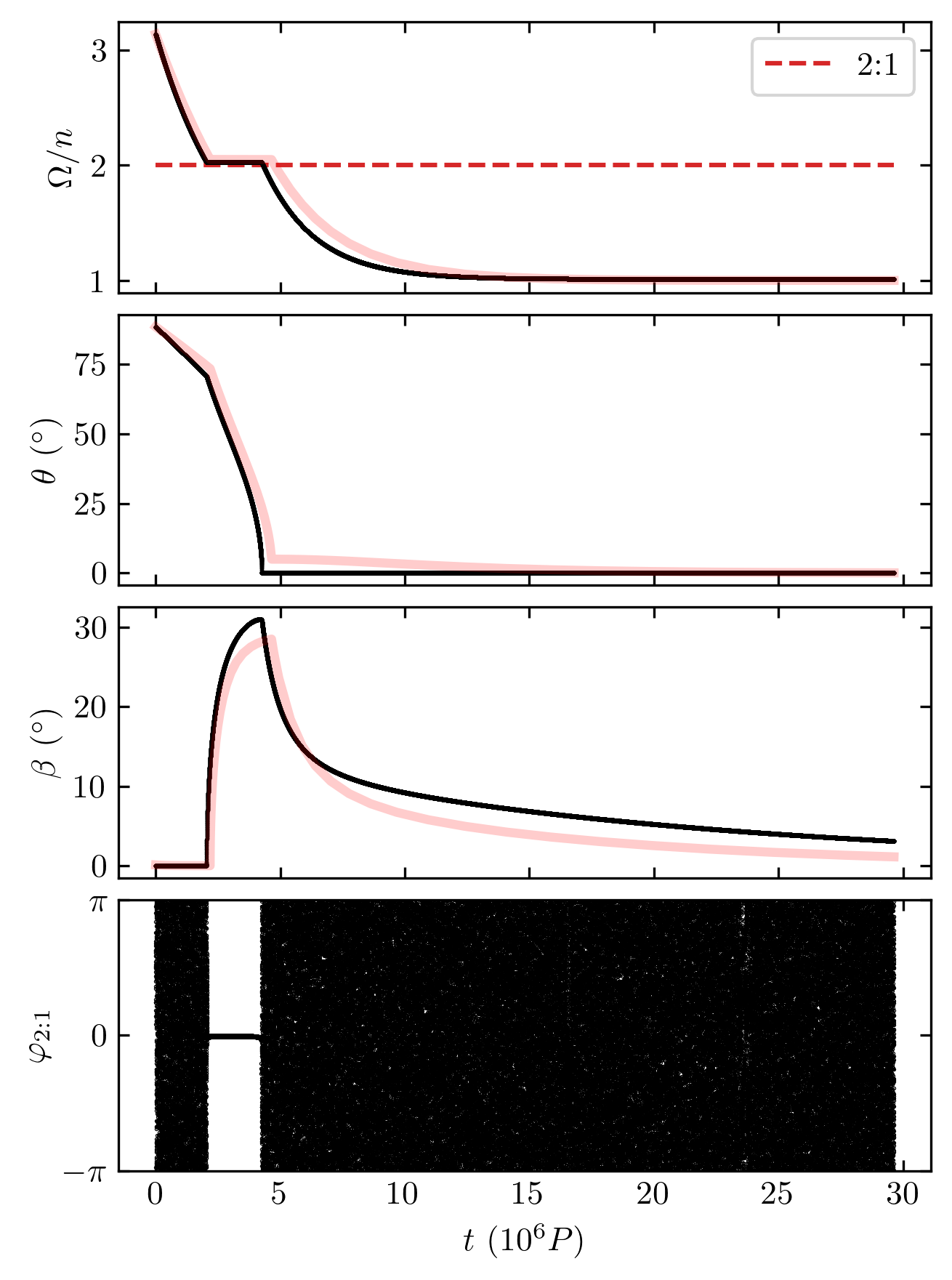}
    \caption{An example integration demonstrating 2:1 resonance capture for an
    oblate planet, where the four panels display the time evolution of the spin
    frequency $\Omega / n$, the obliquity $\theta$, the angle between the
    planet's spin axis and its shortest principal axis $\beta$, and the 2:1
    resonance angle $\varphi_{2:1}$. The initial conditions are $\Omega/n=3.1$
    and $\theta = 88\dgr$. The planet is captured into a 2:1 resonance where
    $\Omega / n \approx 2$ until the obliquity damps to near-zero.
    $\varphi_{2:1}$ is the resonant angle for the 2:1 resonance (see
    Section~\ref{sec:2:1res} for derivation). When the planet is in the
    resonance, this angle librates, and $\beta$ grows substantially while
    $\theta$ damps with increased efficiency. The spin evolution
    through the resonance can be easily understood via the solution presented in
    Section~\ref{ss:npar_evol}, shown by the faint red lines in the top three
    panels.
    }\label{fig:obl-21}
\end{figure}

Next, we discuss the initial conditions of our integrations.
Motivated by the presence of spin-orbit resonances at half-integer values of $\Omega / n$ for eccentric planets, we
choose our initial spin rates to sweep through several small half-integer values of $\Omega / n$
during tidal spindown. We start half of the integrations at $\Omega/n=3.1$ such
that spin-down causes the spin to cross the 3:1, 5:2, and 2:1 resonances (the
last of which the planet is often trapped in), and we start the remaining half
at $\Omega/n=1.6$ such that the spin encounters the 3:2 and 1:1 resonances. We
use 64 initial obliquities evenly spaced from $0\dgr$ to $180\dgr$. Lastly, each
set of initial conditions is run twice---once for a triaxial planet and once for
an oblate planet ($\eta_{\rm tri}=0$)---for a total of 256 integrations.

The results of these integrations are shown in
Figures~\ref{fig:obl-grid}--\ref{fig:triax-grid} for the oblate and triaxial
cases respectively, where several resonance capture features are evident. We
find that oblate and triaxial planets that arrive at $\Omega/n=2$ with
$\theta\lesssim 100\dgr$ are captured into a 2:1 resonance. Once captured, they
continue to evolve with a nearly constant spin rate until their obliquity damps
to near $0\dgr$, at which point their spin rate resumes decreasing towards
synchronous rotation. The dynamics for oblate and triaxial planets are very
similar.

In Fig.~\ref{fig:obl-21}, we show a time-series plot for an
integration consisting of an oblate planet starting from $\Omega / n = 3.1$ with
an initial obliquity of $\theta = 88\dgr$. In the top panel, the spin rate
can be seen to remain near $\Omega / n = 2$ (red dashed
line) for an extended period of time. When the system is in this 2:1 resonance,
the obliquity damping is enhanced (second panel) and non-principal-axis rotation
is excited (third panel). The last panel shows the resonant phase angle for the
2:1 resonance (Section~\ref{sec:ana}), which can indeed be seen to librate when
the system is in resonance.

\section{Non-Dissipative Hamiltonian Resonance Analysis}\label{sec:ana}

In this section, we will develop a Hamiltonian theory for the resonances found
in Section~\ref{sec:num} in the absence of tidal dissipation. We only briefly
summarize the approach and results in the main text, and relegate the
development of the Hamiltonian to Appendix~\ref{app:rotation}.

\subsection{The Hamiltonian and Andoyer Variables}

We begin by expressing the spin Hamiltonian. Here, we will assume that the
planet is on a circular orbit with semi-major axis $a$, and that the spin
angular momentum of the planet is negligible compared to that of the orbit. As
such, we neglect the orbital contribution to the planet's Hamiltonian. Then,
we denote the planet's moment of inertia tensor $\bm{I}$, its spin vector
$\bm{\Omega}$, and its separation vector \emph{from} its host star $\bm{r}$.
With these conventions, the spin Hamiltonian is
\begin{align}
    H &= K + V,\label{eq:H_vec}\\
    K &= \frac{1}{2}\bm{\Omega}^T \cdot \bm{I} \cdot \bm{\Omega},\label{eq:T_vec}\\
    V &= \frac{GM}{2r^5}\s{
            3\bm{r}^T \cdot \bm{I} \cdot \bm{r}
            - \mathrm{Tr}(\bm{I})}\label{eq:V_vec}.
\end{align}
Here, $K$ is the spin kinetic energy, and $V$ is the quadrupolar, leading-order
gravitational potential energy---it is also called MacCulllagh's formula and
leads to the torque given by Eq.~\eqref{eq:rbtriax} \citep{tremaine_book}.

The next step is to express the Hamiltonian in a canonically conjugate set of
coordinates. Although the rotational Euler angles and their conjugate momenta
are the most familiar, a better option for our purposes are the \emph{Andoyer
variables} \citep{andoyer1926cours, peale1973, tremaine_book}. These are a set
of canonical coordinates that are well-suited for studies of spin dynamics,
loosely analogous to Delaunay variables for orbital dynamics\footnote{Other
paths towards a Hamiltonian theory of rotational dynamics use quaternions
\citep[e.g.][]{spring1986euler, udwadia2010alternative, nielsen2012conservative,
2024goldberg} or matrices \citep[e.g.][]{grit}.}.

The Andoyer reference frame is defined by the three orthonormal basis vectors
$\{\uv{X}, \uv{Y}, \uv{Z}\}$ where $\uv{Z}$ is aligned with the planet's spin
angular momentum $\b{S}$, and $\uv{X}$ lies in the space-frame $\uv{x}$-$\uv{y}$
plane. The Andoyer variables consist of the three angles $(l, g, h)$ and their
conjugate momenta $(\Lambda, S, S_z)$. Here, $g$ has the interpretation of being
a pseudo-spin phase of the planet\footnote{We call this angle the spin phase due to its rate of change, but it is more accurately called the phase angle of the free precession (e.g.\ \citealp{landau_lifshitz}) or free nutation (e.g.\ \citealp{tremaine_book}) of the planet in the angular momentum frame.}, $l$ and $h$ are related to the line of nodes
between the Andoyer-body and Andoyer-space frames respectively, and $S =
\abs{\b{S}}$. The last two momenta, given by
\begin{align}
    \Lambda &= S \uv{Z} \cdot {\khat} \equiv S \cos J,\\
    S_z &= S\uv{Z} \cdot \uv{z} \equiv S \cos i,
\end{align}
are the projections of $\b{S}$ along the body and space $z$-axes, respectively.
These three coordinates and conjugate momenta can be shown to be canonically
conjugate \citep[e.g.][]{tremaine_book}. The Andoyer variables
orient the Andoyer frame (the frame aligned with $\b{S}$) with respect to the
space and body frames. A simple pedagogical introduction to Andoyer variables
is given in Appendix~\ref{app:andoyer_def}.

In terms of the Andoyer Variables, the kinetic energy of a rotating body is
given by \citep{tremaine_book}
\begin{align}
    K ={}& \frac{1}{2}\p{\frac{\sin^2l}{A} + \frac{\cos^2 l}{B}}
                \p{S^2 - \Lambda^2} + \frac{\Lambda^2}{2C}.\label{eq:T_andoyer}
\end{align}

\subsection{Oblate body: Hamiltonian Expansion}\label{ss:obl_H}

However, the gravitational potential energy $V$ is also difficult to deal with.
For simplicity, we begin by specializing to the case where the planet is oblate
but axisymmetric, so $A = B$. Eq.~\eqref{eq:H_vec} then can be written
\begin{equation}
    H = \frac{S^2 - \Lambda^2}{2A} + \frac{\Lambda^2}{2C}
        + \frac{3n^2}{2}\s{(C - A)\p{\uv{r} \cdot {\khat}}^2},\label{eq:H_obl}
\end{equation}
where $n = \sqrt{GM/a^3}$ is the planet's mean motion, and we have subtracted
out a constant. It is immediately obvious that for circular orbits, there can
only be $1:1$ and $2:1$ resonances, since the potential only has up to second
harmonics in $\uv{r}$ and ${\khat}$.

It is worth noting that when $n = 0$ (an isolated, axisymmetric
top) that the body satisfies
\begin{align}
    \rd{l}{t} &= \Lambda\p{\frac{1}{C} - \frac{1}{A}},\\
    \rd{g}{t} &= \frac{S}{A}.\label{eq:free_precession}
\end{align}
The first result is just the standard free precession of a symmetric top
\citep[e.g.][]{landau_lifshitz}, and the second result shows that $g \approx
\Omega t$.

To proceed with our analysis of Eq.~\eqref{eq:H_obl}, we next need to express
$\uv{r} \cdot {\khat}$ in terms of the Andoyer variables. For a circular orbit
with mean anomaly $M=nt+\mbox{constant}$, we have
\begin{equation}
    \uv{r} = \cos M \uv{x} + \sin M \uv{y},
\end{equation}
From here, it can be shown that (see Appendix~\ref{app:algebra})
\begin{align}
    \uv{r} \cdot {\khat}
        ={}& \cos\p{M - h}\sin g \sin J - \sin\p{M - h} \nonumber\\
        &\times \p{
                \cos i \cos g \sin J
                + \sin i \cos J}.\label{eq:rdotk_text}
\end{align}

At this point, the essential features of the dynamics become clear, and we
describe them qualitatively before a quantitative analysis. Since $V \propto
(\uv{r} \cdot {\khat})^2$, we immediately see that $V$ consists of a sum of many
trigonometric functions. Two terms of interest are the one with argument $(g -
2M + 2h)$ and the one with argument $(2g - 2M + 2h)$. This suggests that
resonances can occur when $2\dot{M} = \dot{g}$ or when $\dot{M} = \dot{g}$
(since $\dot{h} \sim \mathcal{O}(C - A)$ is negligible). Since we showed that
$\dot{g} \approx \Omega$ above, this shows that resonances can occur when
$\Omega \approx 2n$ or when $\Omega \approx n$. 
The former
resonance is evident in our numerical results of Section~\ref{sec:num}, while
the latter is suppressed by NPAR damping (see later discussion in
Section~\ref{sec:2:1res}). The fact that bodies undergoing non-principal-axis rotation experience a
component of the perturbing potential involving the second harmonic of the mean
motion (i.e.\ terms like $\sin 2M$) has been pointed out in previous works
\citep[e.g.][]{peale1973, efroimsky2001}, but the possibility of resonant
dynamics involving this harmonic has not been previously explored.

Note that for bodies undergoing principal axis rotation ($J = \sin J = 0$),
Eq.~\eqref{eq:rdotk_text} no longer depends on $g$, and thus these resonances
disappear. This is due to our assumption of axisymmetry, since the torque
experienced by a zero-obliquity, axisymmetric body does not change as the body
spins.

\subsection{The 2:1 Resonance for an Oblate Planet}\label{sec:2:1res}

We now analyze these resonances more quantitatively. After a bit of algebra (see
Appendix~\ref{app:algebra}), we find that the term in the expansion of $(\uv{r}
\cdot {\khat})^2$ that is relevant to the 2:1 resonance is
\begin{align}
    \p{\uv{r} \cdot {\khat}}^2_{2:1}
        ={}& -\p{1 + \cos i}\nonumber\\
            &\times \frac{\sin 2J\sin i}{4} \cos\p{g - 2M + 2h}.
                \label{eq:rdotk_21}
\end{align}
This results in the 2:1 resonant Hamiltonian
\begin{align}
    H_{2:1} \approx{}& \frac{S^2 - S^2\cos^2J}{2A}
                + \frac{S^2\cos^2J}{2C}\nonumber\\
            &- \frac{3n^2(C - A)}{8}\sin i (1 + \cos i)\nonumber\\
            &\times\sin 2J\cos\p{g - 2M + 2h}.
\end{align}
It is worth noting that $l$ is ignorable, so $\Lambda$ is conserved exactly.
The evolution of $g$ is dominated by the free precession solution
(Eq.~\ref{eq:free_precession}). Comparatively, since $(C - A) / A \sim J_2 \ll
1$, the evolutions of $h$, $S$, and $S_z$ are much slower.

We next perform a canonical transformation such that the desired resonant angle,
given by
\begin{equation}
    \varphi_{2:1} = g - 2nt + 2h
\end{equation}
is one of the coordinates. The bottom panel of Fig.~\ref{fig:obl-21} shows that
this angle is indeed librating when the planet is captured in the 2:1 resonance
in our numerical integrations. By constructing the type-2 generating function
\citep{landau_lifshitz}
\begin{equation}
    F_2\p{g, h, S, S_z', t} = S\p{g - 2nt + 2h}  + S_z'h,
\end{equation}
we see that $S_z$ must be replaced by $S_z' \equiv S_z - 2S$. The resulting
Hamiltonian is then:
\begin{align}
    H\p{l, \varphi_{2:1}, h; \Lambda, S, S_z'}
        ={}& \frac{S^2 - \Lambda^2}{2A} + \frac{\Lambda^2}{2C}
            - 2nS \nonumber\\
            &- \frac{3n^2(C - A)\cos\p{\varphi_{2:1}}}{4}\nonumber\\
            &\times \frac{\sqrt{S^2 - (S_z' + 2S)^2}}{S^4}\nonumber\\
            &\times (S_z' + 3S)\sqrt{S^2 - \Lambda^2}\Lambda.\label{eq:H_21}
\end{align}
We have explicitly shown all of the actions and angles as arguments to the
resonant Hamiltonian for clarity. Note that $l$ and $h$ do not appear on the
right side of eq.~\eqref{eq:H_21}, so that $\Lambda$ and $S'_z$ become constants
in this approximation. Hamilton's equations then give for the resonant angle and
momentum pair:
\begin{align}
    \rd{\varphi_{2:1}}{t} = \pd{H}{S}
        ={}& \p{\frac{S}{A} - 2n} + \mathcal{O}\p{C - A},\\
    \rd{S}{t} = -\pd{H}{\varphi_{2:1}}
            ={}& -\frac{3n^2(C - A)\sin\p{\varphi_{2:1}}}{4}\nonumber\\
                &\times \frac{\sqrt{S^2 - (S_z' + 2S)^2}}{S^4}\nonumber\\
                &\times (S_z' + 3S)\sqrt{S^2 - \Lambda^2}\Lambda.
\end{align}
Indeed, this is exactly the resonant behavior we found in Section~\ref{sec:num},
and suggests that when $S / A \approx 2n$ that a resonance occurs.

A useful result is to compute the resonant libration frequency $\omega_{\rm lib,
2:1}$, defined by
\begin{align}
    \rtd{{\varphi}_{2:1}}{t} &= \dot{S} / A\nonumber\\
        &\equiv -\omega_{\rm lib, 2:1}^2\sin \varphi_{2:1}.
\end{align}
We approximate $S \approx 2nA$ and obtain
\begin{align}
    \omega_{\rm lib, 2:1}^2 ={}&
        \frac{3n^2(C - A)}{4A}
            \frac{\sqrt{4n^2A^2 - \p{S_z' + 4nA}^2}}{16n^4A^4}\nonumber\\
        &\times \p{S_z' + 6nA}\sqrt{4n^2A^2 - \Lambda^2}\Lambda,\nonumber\\
        \approx{}& \frac{3n^2(C - A)}{8A}
            \sin i \sin 2J \p{1 + \cos i}.\label{eq:wlib_21}
\end{align}
In the last line, we have also approximated $\Lambda = S \cos J \approx 2nA\cos
J$ (and correspondingly for $S_z'$), which is accurate for small libration
amplitudes and large angles $J$ and $i$---since $\Lambda$ is conserved while $S$
varies during resonant libration, the approximate Eq.~\eqref{eq:wlib_21} only
holds when $S^2 - \Lambda^2 \propto S^2\sin^2J$ is much greater than the
variation of $S^2$.

Note that an analogous procedure can be used to obtain a 1:1
resonance feature, with resonant angle $(2g - 2nt + 2h)$. The width of this
resonance scales like $\sin^2J$ though (Eq.~\ref{eq:rdotk_full}), and is
overwhelmed by the NPAR damping torque for the parameters we adopt.

Lastly, we note that it is also of course possible to repeat the above procedure
starting from Eqs.~(\ref{eq:H_vec},~\ref{eq:T_andoyer}) without assuming that $B
= A$. The resulting expressions are considerably more complex owing to the
explicit appearance of $l$, which leads to non-conservation of $\Lambda$; the
full expansions are available in a Sympy
notebook\footnote{https://github.com/yubo56/ipynbs/blob/main/Andoyer\_check.ipynb}.
In broad summary, we find that the coefficients of the relevant 2:1 resonant
terms responsible for Eqs.~\eqref{eq:H_21} do not change, but many more resonant
angles become possible due to additional $l$ dependencies.

\subsection{Description of Resonant Spin Evolution}\label{ss:npar_evol}

While it is in principle possible to modify the equations of motion for the
resonance to include the equilibrium tidal dissipation described by
Eq.~\eqref{eq:tidaltorque}, the behaviors observed in
Figs.~\ref{fig:obl-grid}--\ref{fig:obl-21} can be understood without significant
additional algebra. To simplify the discussion below, we will assume that
$\eta_{\rm obl}$ is sufficiently small such that $S \propto \Omega$, nearly
independent of the spin-body misalignment angle $J$, so that the angles $i
\approx \theta$ and $J \approx \beta$ are used interchangeably\footnote{Recall that $\theta$ and $\beta$ are the misalignment angles of the spin vector $\uv{\Omega}$ to the orbit normal and the short body axis respectively, while $i$ and $J$ are those of the spin \emph{angular momentum} $\bm{S}$ to $\uv{z}$ and $\khat$. The differences between these angles are $\sim \mathcal{O}(\eta_{\rm obl})$ and are thus small.}.

The spin evolution evolves under the combined effect of tidal dissipation and
the rigid body dynamics. The latter is governed by the resonant Hamiltonian
(Eq.~\ref{eq:H_21}). Since the angles $l$ and $h$ are absent from the
Hamiltonian, their conjugate actions $\Lambda = S\cos J$ and $S_z' = S \cos i -
2S$ do not change due to the rigid body dynamics alone. On the other hand, when
the system is trapped in resonance, $S$ and $\Omega$ are fixed. These
observations let us write out, when the system is in resonance:
\begin{align}
    \rd{S}{t}
        &= \p{\rd{S}{t}}_{\rm tide}
            + \p{\rd{S}{t}}_{\rm RB} = 0,\label{eq:dsdt_rb}\\
    \rd{\cos i}{t}
        &= \p{\rd{\cos i}{t}}_{\rm tide}
            + \p{\rd{\cos i}{t}}_{\rm RB}\nonumber\\
        &= \p{\rd{\cos i}{t}}_{\rm tide}
            - \frac{2 - \cos i}{S}\p{\rd{S}{t}}_{\rm tide},\\
    \rd{\cos J}{t}
        &=  \frac{\cos J}{S}\p{\rd{S}{t}}_{\rm tide}.\label{eq:djdt_rb}
\end{align}
Using the standard expressions for the constant time lag tidal model
(consistent with our Eq.~\ref{eq:tidaltorque}, \citealp{hut1981tidal, 2012Lai})
\begin{align}
    \frac{1}{S}\p{\rd{S}{t}}_{\rm tide}
        &= \frac{1}{t_{\rm tide}}\s{\frac{n}{\Omega}\cos i - \frac{1 +
            \cos^2i}{2}},\label{eq:dsdt_tide}\\
    \p{\rd{i}{t}}_{\rm tide}
        &= -\frac{\sin i}{t_{\rm tide}}\s{\frac{n}{\Omega} - \frac{\cos i}{2}},
            \label{eq:didt_tide}
\end{align}
where $t_{\rm tide}$ is given by Eq.~\eqref{eq:tau_tide}, the system evolution
in resonance can easily be described. Combining the out-of-resonance
(Eqs.~\ref{eq:dsdt_tide}--\ref{eq:didt_tide}, and NPAR damping given by
Eq.~\ref{eq:tau_npar}) and in-resonance (inclusion of
Eqs.~\ref{eq:dsdt_rb}--\ref{eq:djdt_rb}) evolution,
and assuming that resonance ejection occurs when $i$ is sufficiently small
(e.g.\ $\lesssim 5^\circ$), we can approximate the full spin evolution of the
planet. Note that our expressions do not depend on
$\varphi_{2:1}$ and so can easily be used in a secular-averaged setting. This is
shown as the thick red lines in Fig.~\ref{fig:obl-21}, where good agreement is
obtained.



\subsection{Resonance Capture}\label{ss:adiabat}

In this section, we briefly comment on the resonance capture conditions. It is
well-known that resonance capture requires that the passage be adiabatic, such
that the resonance is crossed much slower than the characteristic libration
frequencies of the system \citep[e.g.][]{ward-hamilton2004-Ianalytic, su2020}.
For the dynamics we consider, the resonance crossing time is set by $\rdil{\ln
\Omega}{t}$ (Eq.~\ref{eq:dsdt_tide}), and the libration frequency of the system
is given by Eq.~\eqref{eq:wlib_21}. However, it is not straightforward to apply
Eq.~\eqref{eq:wlib_21}: it vanishes for $J = 0$, while the planet starts out in
principal-axis rotation.

Instead, we note that, in the vicinity of the resonance, $S$ and therefore $J$
begin to oscillate. By evaluating the resonant Hamiltonian (Eq.~\ref{eq:H_21})
at $\varphi_{2:1} = 0$ and $\varphi_{2:1} = \pi$, the amplitude of the
oscillations in $J$ during resonance approach can be estimated:
\begin{equation}
    \Delta J \sim \frac{3\sqrt{2}n^2(C - A)A}{4S^2}
        \sin i(1 + \cos i).
\end{equation}
From this, we can evaluate $\omega_{\rm lib, 2:1}$ for $J \sim \Delta J / 2$,
which yields
\begin{equation}
    \omega_{\rm lib, 2:1} \approx \frac{3}{4\sqrt{2}}
        \eta_{\rm obl}\frac{n}{\Omega} \sin i\p{1 + \cos i}n.
\end{equation}
Thus, the adiabaticity can be evaluated:
\begin{align}
    \mathcal{A}
        \equiv{}& \omega_{\rm lib, 2:1}\abs{\rd{\ln \Omega}{t}}^{-1},\\
        ={}& \frac{\eta_{\rm obl}}{4\sqrt{2}}
            \frac{Q}{k_2}
            \frac{m}{M_\star}
            \p{\frac{a}{R}}^3
            \frac{2\sin i(1 + \cos i)}{1 + \cos^2 i - \cos i}\nonumber\\
        \approx{}& 35
            \frac{2\sin i(1 + \cos i)}{1 + \cos^2 i - \cos i}
            \p{\frac{\eta_{\rm obl}}{10^{-5}}}
            \p{\frac{k_2/Q}{10^{-3}}}^{-1}\nonumber\\
            &\times \p{\frac{m}{1.57 M_{\oplus}}}
            \p{\frac{M_\star}{1.2M_{\odot}}}^{-1}\nonumber\\
            &\times \p{\frac{a}{0.124\;\mathrm{AU}}}^3
            \p{\frac{R}{1.176 M_{\oplus}}}^{-3}.\label{eq:A_cond}
\end{align}
Typically, $\mathcal{A} \gg 1$ is sufficient to ensure adiabatic passage, and
therefore resonance capture.

\begin{figure*}
    \centering
    \includegraphics[width=0.7\textwidth]{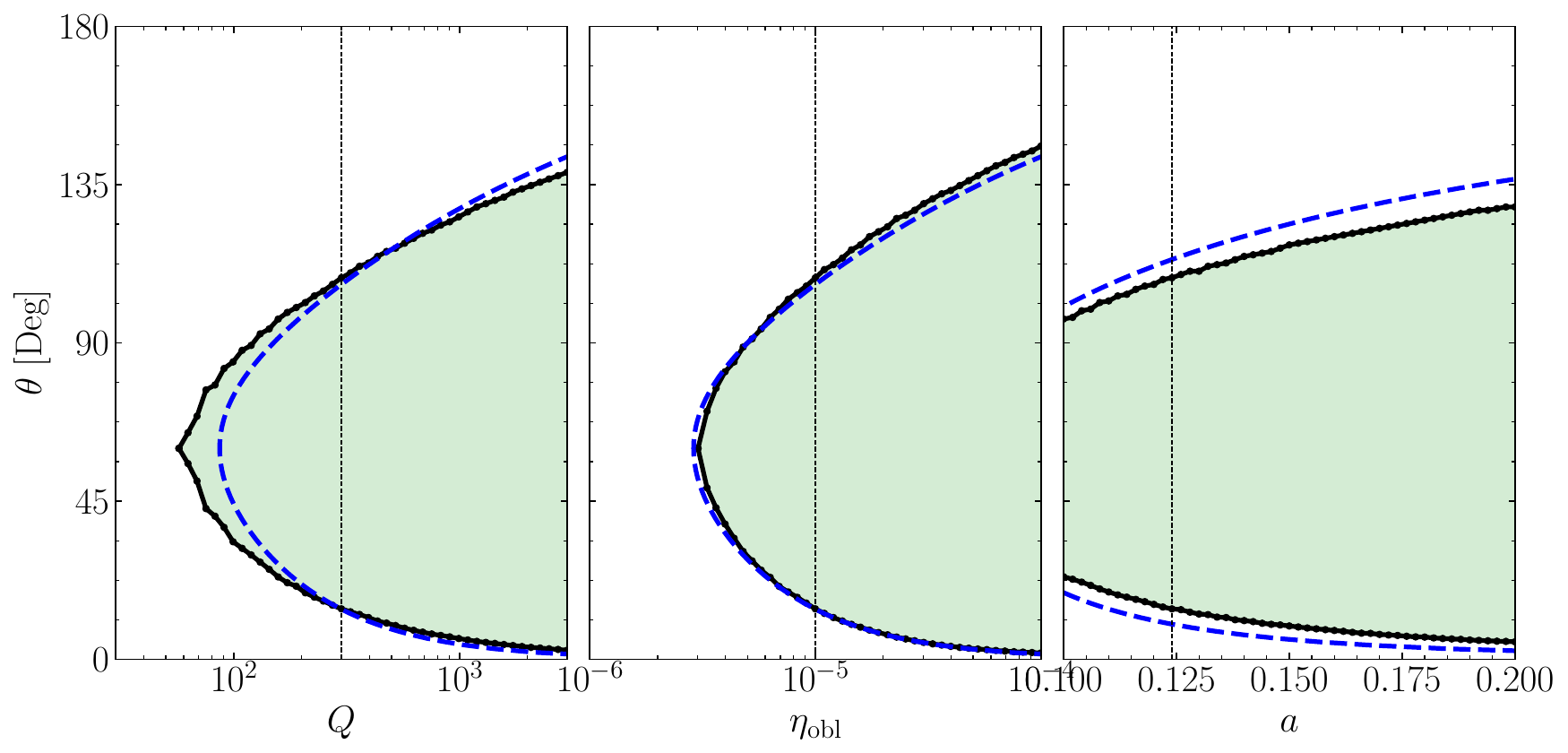}
    \caption{Region of $\theta$ values (green) that undergo capture into the 2:1
    resonance when varying the tidal $Q$ of the planet, the $\eta_{\rm obl} =
    J_2/k$ of the planet, and the planet's semi-major axis $a$.
    The fiducial values of all three parameters are marked with
    the black dashed lines. It can be seen that stronger tidal dissipation
    (decreased $Q$ or decreased $a$) or a weaker resonance (decreased $J_2$)
    reduce or even eliminate the obliquity range experiencing resonance capture.
    The blue dashed curve identifies the condition that the adiabaticity, given
    by Eq.~\eqref{eq:A_cond}, satisfies $\mathcal{A} \gtrsim 35$.
    }\label{fig:3cutoffs}

\end{figure*}

In Figure~\ref{fig:3cutoffs}, the green shaded region indicates
the parameter space where resonance capture occurs, where we vary the tidal
dissipation strength (via $a$ and $Q$) and the resonance strength (via
$\eta_{\rm obl}$). The blue dashed line in all three panels identifies the curve
along which $\mathcal{A} = 35$. Thus, we conclude that $\mathcal{A} \gtrsim 35$
results in resonance capture. The fact that the critical $\mathcal{A}$ is not of
order unity likely reflects the detailed geometry of the resonance capture
process (e.g.\ a longer libration time than $\omega_{\rm lib, 2:1}$ due to
evolution near the separatrix). Finally, note as well that, when $\eta_{\rm obl}
\lesssim 3 \times 10^{-6}$, resonance capture does not occur (middle panel of
Fig.~\ref{fig:3cutoffs}). In conjunction with our estimated maximal asphericity
from Eq.~\eqref{eq:rigidity}, this suggests that higher-mass SEs may avoid
resonance capture, unless their $Q$s are larger than the $Q = 300$ we've assumed
here.


\section{Summary \& Discussion}\label{sec:disc}

In this paper, we study the spin evolution of a single rigid super Earth (SE) on
a circular orbit undergoing tidal dissipation. We assume that the planet is
formed spinning mildly supersynchronously but make no assumptions on its initial
obliquity (denoted $\theta$). Denoting the planet's mean motion by $n$, our
primary results are that:
\begin{itemize}
    \item The planet's tidal spindown can often be temporarily arrested when the
        planet's spin rate $\Omega$ satisfies $\Omega =
        2n$, but not other values (see Section~\ref{sec:num} and
        Figs~\ref{fig:obl-grid} and~\ref{fig:triax-grid}).

    \item Capture into these non-secular spin-orbit resonances occurs for a
        broad range of $\theta$ and succeeds as long as the resonance is encountered adiabatically (Eq.~\ref{eq:A_cond} and Section~\ref{ss:adiabat}).

    \item Upon capture into either resonance, $\theta$ begins to damp at an
        accelerated rate while $\Omega$ remains approximately constant until
        $\theta$ decreases below $\sim 5$--$10\dgr$ and the planet exits the
        resonance. While the planet is in the resonance, tidal dissipation acts
        to increase the deviation of the planet from principal-axis rotation
        (see Fig.~\ref{fig:obl-21} and Section~\ref{ss:npar_evol}).
\end{itemize}
We show in Section~\ref{sec:ana} that these resonances can be understood
analytically, and we derive the resonant
angle, approximate tidal evolution in resonance, and the resonance capture
condition.

\subsection{Discussion: Exoplanetary Obliquities}

Spin-orbit resonances like these have been previously reported as
``inclination-driven resonances'' for planets obeying a Maxwell viscoelastic
rheology \citep{boue-etal2016} and for planets
obeying an Andrade, anelastic rheology \citep[suitable for rocky
planets,][]{revol-etal2023}. Both works reproduce almost all of the dynamical
features of our resonances: planets on circular orbits experiencing tidal
dissipation are trapped at $\Omega = 2n$ (though they also find capture at $\Omega = n$ and $\Omega = 0$) for an intermediate range of obliquities and are ejected
from the resonance when the obliquity becomes sufficiently small due to tidal
dissipation.
Despite this apparent similarity, their results arise due to their choice of rheology and thus are physically distinct from ours:
in their work, the resonant feature arises from a vanishing of the
secular-averaged tidal despinning rate when (i) one tidal component (i.e.\ one
of the tidal frequencies $\equiv mn - m'\Omega$) has a much lower frequency than the others, and (ii) when the orbital period is much shorter than some relaxation
timescale of the planet \citep{dobrovolskis1980, correia2022formalism}. However,
estimates of this relaxation timescale vary over several orders of magnitude
and depend on the planet's properties \citep[e.g.][]{storch2014visco,
correia2019}, so this condition may not be satisfied for all rocky exoplanets.
Even if it is satisfied, the tidal despinning of a planet may not arise from
dissipation in the bulk of the body: for instance, dissipation in the Earth-moon
system is dominated by that in the Earth's oceans \citep[e.g.][]{egbert2000,
ray2001}.
Broadly speaking, the results in the current paper suggest that adoption of a rheology where the secular tidal despinning rate vanishes at spin-orbit commensurabilities is \emph{not} required for a planet to
be trapped in the $\Omega = 2n$ resonance, as long as NPAR rotation
is allowed.
As such, the prevalence of the 2:1 spin-orbit resonance for rocky bodies may be
more general than previously thought.

We also briefly comment on the effect of these resonances on the conclusions of
\citet{su-lai2022-res}, which concluded that SEs with cold Jupiter companions
are trapped in a high-obliquity secular spin-orbit resonance (``Cassini State
2''; CS2) $\sim 30\%$ of the time. The high-obliquity resonances studied in
their work occur for $\Omega < n$ and in fluid planets, where there are no
non-secular spin-orbit resonances. \citet{gladman1996} studied the analogous
spin-orbit dynamics for rocky satellites. They find that two possible CS2
solutions can exist for rocky planets, one at subsynchronous rotation and one at
synchronous rotation. They find that the former generally only exists at low
obliquities, and the latter is unstable when $\theta \gtrsim 68^\circ$. As such,
no rocky bodies can retain large obliquities by being trapped in CS2, in
agreement with observations of solar system satellites \citep{peale1977} and in
contrast with the results of \citet{su-lai2022-res} for fluid-like planets.
However, the allowance of non-PAR in this work introduces a second possible CS2
solution at the $\Omega = 2n$ resonance, which may be tidally stable. Such
resonances may give more ways for close-in exoplanets to avoid becoming tidally
locked. In addition, the rotation rates of high-obliquity
exoplanets may reflect their physical properties, with subsynchronous rotators
likely being fluid-like and supersynchronous ones likely being rocky. Further
study of this possibility and its effect on the obliquities of SEs will be
considered in future work.

One caveat of this work is that it neglects the dynamical effect of exomoons.
Earth's spin evolution is substantially complicated by the formation of the moon
and the evolution of its orbit \citep[e.g.][]{laskar1993_moon, touma1998,
lissauer2012, li2014moon, rufu2020}. In addition, migrating satellites have been
proposed to affect the obliquities of Jupiter, Saturn, and Uranus
\citep{saillenfest-etal2020, saillenfest-etal2021-migration,
saillenfest-etal2022, wisdom2022}. While the search for exomoons is ongoing
\citep{teachey2018moon, kipping2022}, their presence can affect obliquity
evolution \citep{saillenfest2023}. We plan to study these dynamics in future
work.

Another caveat of our work is the simple form of the non-principal-axis rotation
(NPAR) damping adopted in this work. First, more sophisticated analyses of
non-PAR damping suggest that there are many more efficient mechanisms for
non-PAR damping in certain regimes \citep{Yoder1979, efroimsky2001}.
However, given the observation of the Chandler wobble over a
$\sim 100\;\mathrm{yr}$ baseline, as well as evidence for NPAR rotation in
Venus \citep{yoder1979venuswander, spada1996venuswander} and in Mars
\citep{schultz1988}, NPAR damping is likely not significantly more efficient
for an Earth-like planet than Eq.~\ref{eq:tau_npar} predicts. Future prospects
for constraining NPAR damping efficiency may be obtained from the well-known
tumbling of Hyperion \citep{1984wisdom, wisdom1987, goldberg2024}. Second, in
the last line of Eq.~\eqref{eq:tau_npar}, we have assumed that $Q$ is comparable
for the two different forcing frequencies, the tidal ($n$) and wobble
($\Omega$). Of course, in general, $Q$ is not constant and depends on the
forcing frequency (as the constant-phase-lag-model is nonphysical, e.g.\
\citealp{lu-etal2023}). However, since $\Omega \sim n$, our assumption that the
$Q$ values are comparable is likely good unless $Q$ does not vary smoothly as a
function of frequency (which may be the case at much faster frequencies due to
planetary oscillation modes). Finally, the non-PAR damping rate adopted here is
due to damping of the internal stresses generated by the time-varying
centrifugal potential. A second contribution to non-PAR damping can arise if the
planet flows or deforms in response to the gravitational potential
\citep[e.g.][]{correia-etal2014}. We neglect this contribution here, but it is
of course the dominant channel for fluid-like rheologies (e.g.\ planets with
extended fluid envelopes).

\section*{acknowledgements}

We thank Joshua Winn for insightful comments that improved the quality of this
manuscript. YS thanks Gwena\"el Bou\'e, Brett Gladman, Max Goldberg, Sam Hadden,
Yoram Lithwick, Tiger Lu, Sarah Millholland, Phil Nicholson, Pierre-Louis Phan,
Alexandre Revol, and Melaine Saillenfest for useful discussions and
comments, and YS thanks Dong Lai for the first suggestions for this work.
YS is supported by a Lyman Spitzer, Jr. Postdoctoral Fellowship at Princeton
University.

\software{
GRIT \citep{grit}, Jupyter \citep{jupyter},
matplotlib \citep{matplotlib}, numpy \citep{numpy}, REBOUND \citep{rebound},
REBOUNDx \citep{reboundx}, scipy \citep{scipy}, sympy \citep{sympy}}

\appendix

\section{Rotational Dynamics}\label{app:rotation}

We include here some general background on our notations and approach to the
analytical results on rotational dynamics given in Section~\ref{sec:ana}.

\subsection{Notations and Convention for Rotation}

We begin by defining some notations. As in the main text, we denote the inertial
space-frame basis vectors by $\{\xhat, \yhat, \hat{\bm{z}}\}$ and
the body-frame basis vectors by $\{\ihat, \jhat,
\hat{\bm{k}}\}$. When necessary, we will denote the components of a vector
$\bm{v}$ in the space frame by $\bm{v}_s$ and in the body frame by $\bm{v}_b$.
Using this notation, we define the change-of-basis matrix $\bm{R}_{sb}$ such
that
\begin{equation}
    \bm{v}_s = \bm{R}_{sb} \bm{v}_b.
\end{equation}
Note that the matrix $\bm{R}_{sb}$ also encodes the orientation of the body via
three \emph{Euler angles} \citep[e.g.][]{landau_lifshitz}, which we denote as
$\phi$, $\theta$, and $\psi$. We adopt the ZXZ convention for the order of the
Euler angle rotations (see Eq.~\ref{eq:eulermat}), and notate Euler rotation
matrices as
\begin{equation}
    \bm{R}_{sb} = \bm{R} (\phi, \theta, \psi).
\end{equation}
The components of the Euler rotation matrix are given (we adopt the active
transformation convention)
\begin{align}
    \bm{R}\p{\phi, \theta, \psi} =
    \begin{bmatrix}
        \cos \phi \cos \psi - \cos\theta \sin\phi \sin\psi &
        -\cos\phi\sin\psi - \cos\theta \cos\psi \sin\phi &
            \sin\phi \sin\theta\\
        \cos\psi \sin\phi + \cos\phi \cos\theta \sin\psi &
        \cos\phi \cos\theta \cos\psi - \sin\phi \sin\psi &
            -\cos\phi \sin\theta\\
        \sin\theta \sin\psi &
        \cos\psi \sin\theta &
        \cos\theta
    \end{bmatrix}\label{eq:eulermat}
\end{align}

\subsection{Andoyer Variables}\label{app:andoyer_def}

We expand on our introduction of the Andoyer variables and our notations here;
for a pedagogical treatment, please see Section~7.3 of \citet{tremaine_book}.
First, let $\{\uv{X}, \uv{Y}, \uv{Z}\}$ be the orthonormal basis vectors of the
Andoyer reference frame, and denote the components of a vector $\bm{v}$ in the
Andoyer frame by $\bm{v}_A$. The Andoyer frame is defined as the frame where
$\uv{Z} \propto \bm{S}$ the spin angular momentum (AM), and $\uv{X}$ is in the
$\uv{x}$--$\uv{y}$ plane (orbital plane). The Andoyer variables are obtained by
considering the Euler angles describing the rotations from the Andoyer frame to
the space and body frames. In particular, define the two Euler angles $h$ and
$i$ such that
\begin{align}
    \bm{R}_{sA} &\equiv \bm{R}\p{h, i, 0}.
\end{align}
Recalling that $\bm{R}_{sb} = \bm{R}\p{\phi, \theta, \psi}$, we can write
\begin{align}
    \bm{R}_{Ab} &= \bm{R}\p{0, -i, -h}\bm{R}\p{\phi, \theta, \psi}\nonumber\\
        &\equiv \bm{R}\p{g, J, l},\label{eq:R_sAb}
\end{align}
where $g$, $J$, and $l$ are three new Euler angles. It can then be shown that
the angles $\p{l, g, h}$ are canonically conjugate to the following momenta
\citep{tremaine_book}:
\begin{align}
    p_l &= \bm{S} \cdot {\khat} \equiv \Lambda = S\cos J,\nonumber\\
    p_g &= S,\nonumber\\
    p_h &= \bm{S} \cdot \uv{z} \equiv S_z = S\cos i.\label{eq:andoyer_conjs}
\end{align}
where $S$ is the magnitude of the spin AM, and we have defined $J$ and $i$ to be
the misalignment angles of the spin AM with respect to the body ${\khat}$ axis
and the spatial $\uv{z}$ axis.

\subsubsection{Translation between Vectors and Andoyer Variables}

We include here a brief procedure for converting between the spin states that
are tracked by our code and the Andoyer variables. Our code represents the state
of the planet with four space-frame vectors $\uv{r}_s$,
$\ihat_s$, $\jhat_s$, and $\khat_s$ (we have notated the coordinate basis
in which each vector is stored; we of course also track the total spin rate
$\Omega$) as well as the body-frame spin vector $\uv{\Omega}_b$.
With these, we can construct the change-of-basis matrices:
\begin{itemize}
    \item $\bm{R}_{sb} = \s{\ihat_s^T, \jhat_s^T, \khat_s^T}$, where
        the columns contain the space-frame coordinates of the body-frame basis vectors.

    \item Next, we seek $\bm{R}_{sA}$, for which we need to compute the Andoyer
        basis vectors $\z{\uv{X}, \uv{Y}, \uv{Z}}$. $\uv{Z}$ is located along
        the spin AM $\bm{S} \equiv \bm{I} \cdot \bm{\Omega}$, $\uv{X}$ is
        located along $\uv{z} \times \uv{Z}$, and $\uv{Y} = \uv{Z} \times
        \uv{X}$. Then $\bm{R}_{sA} = \s{\uv{X}_s^T, \uv{Y}_s^T, \uv{Z}_s^T}$.

        Note that $\bm{R}_{sA} = \bm{R}\p{h, i, 0}$.

    \item Finally, we compose these matrices to obtain $\bm{R}_{Ab} =
        \bm{R}_{As}\bm{R}_{sb}$.

        Note that $\bm{R}_{Ab} = \bm{R}\p{g, J, l}$.
\end{itemize}
From the components of $\bm{R}_{sA}$ and $\bm{R}_{Ab}$, we can solve
Eq.~\eqref{eq:eulermat} to obtain the values of $h$, $i$, $g$, $J$, $l$, and
then using the value of $S$ (the spin AM magnitude), we can compute all six
Andoyer variables.

\section{Gravitational Potential Energy in Andoyer Variables}\label{app:algebra}

\subsection{Oblate Planet}

In Section~\ref{ss:obl_H}, we found that the non-Keplerian component of the
gravitational potential energy for an oblate body in a gravitational field is
given up to an additive constant by
\begin{equation}
    V = \frac{3n^2}{2}\s{(C - A)\p{\uv{r} \cdot {\khat}}^2}\label{eq:V_obl}.
\end{equation}
This subsection is dedicated to the expansion of this expression in Andoyer
variables.

First, we note that
\begin{align}
    \uv{r}_s ={}& \begin{bmatrix}
        \cos M \\
        \sin M \\
        0
    \end{bmatrix},\\
    {\khat}_s ={}& \bm{R}_{\rm sb} \begin{bmatrix}
        0\\0\\1
    \end{bmatrix} =
    \bm{R}\p{h, i, 0}\bm{R}\p{g, J, l}\begin{bmatrix}
        0\\0\\1
    \end{bmatrix},\\
    \uv{r}_s \cdot {\khat}_s
        ={}& \cos\p{M - h}\sin g \sin J\nonumber\\
            &- \sin\p{M - h}\p{
                \cos i \cos g \sin J
                + \sin i \cos J}.
\end{align}
This is Eq.~\eqref{eq:rdotk_text}, where of course the dot product does not
depend on the basis in which it is evaluated.

Then, expanding $(\uv{r} \cdot {\khat})^2$, we can group the resulting terms as
follows (for simplicity, denote $M' \equiv M - h$):
\begin{align}
    4\p{\uv{r} \cdot {\khat}}^2 ={}& \Bigg\{
            \sin^2J\p{1 - \cos 2g}
            + \cos^2i\p{1 + \cos 2g}\sin^2J
            + \sin 2i \sin 2J \cos g
            + 2\sin^2i \cos^2J
        \Bigg\}_{\rm (i)}\nonumber\\
        &+ \z{
            \cos 2M' \sin^2J
            - \cos 2M' \cos^2 i \sin^2 J
            - 2\cos 2M'\sin^2 i \cos^2 J
        }_{\rm (ii)}\nonumber\\
        &+ \Bigg\{
            -\cos 2M' \sin 2i \sin 2J \cos g
            - 2\sin 2M' \sin 2J \sin i \sin g
        \Bigg\}_{\rm (iii)}\nonumber\\
        &+ \Bigg\{
            -\cos 2M' \cos 2g \sin^2 J
            - \cos 2M' \cos^2 i \cos 2g \sin^2J
            - 2\sin 2M' \sin 2g \sin^2 J \cos i
        \Bigg\}_{\rm (iv)}.\label{eq:rdotk_full}
\end{align}
Here, the four curly-bracketed terms split the expression into: (i) no $M'$
dependence, (ii) $M'$ dependence but none on $g$, (iii) depends on $\varphi_{2:1} =
2M' \pm g$, and (iv) depends on $\varphi_{1:1} = 2M' \pm 2g$.

When attempting to average Eq.~\eqref{eq:rdotk_full} over time scales $\gg
n^{-1}, \Omega^{-1}$, there are three cases: $n\approx \Omega$, $2n \approx
\Omega$, and neither. In all three cases, expression (ii) vanishes while
expression (i) persists. The terms in expression (i) can be shown to
straightforwardly reduce to the standard Colombo's Top when assuming principal
axis rotation \citep[e.g.][]{tremaine_book}. The remaining expressions (iii) and
(iv) reduce to Eq.~\eqref{eq:rdotk_21} when $2n \approx \Omega$. There are also
resonant terms with resonant angle $2M' + g$ and $2M' + 2g$, but these only
appear for negative spin rates and are equivalent to a sign flip in $\cos i$.

\subsection{Triaxial Planet}

A similar procedure to the previous section can be performed for a
non-axisymmetric, triaxial body satisfying $A < B < C$. In this case, the
potential given by Eq.~\eqref{eq:V_vec} can be written up to an additive
constant as
\begin{equation}
    V = \frac{3n^2}{2}\s{(C - A)\p{\uv{r} \cdot {\khat}}^2
        + (B - A)\p{\uv{r} \cdot {\jhat}}^2}.
\end{equation}
This can be evaluated by expressing ${\jhat}$ in Andoyer variables in space-frame coordinates using
\begin{align}
    {\jhat}_s ={}& \bm{R}\p{h, i, 0}\bm{R}\p{g, J, l}\begin{bmatrix}
        0\\1\\0
    \end{bmatrix}.
\end{align}
The resulting expansion is laborious and mostly uninsightful, and it can be
performed with computer
algebra\footnote{https://github.com/yubo56/ipynbs/blob/main/Andoyer\_check.ipynb}.
The character of the $(g - 2M)$ and $(2g - 2M)$ resonances do not change, though
new resonant angles can appear with dependencies on the Andoyer angle $l$.

\section{Principal Axis Rotation Damping}\label{app:goodman_damping}

In this section, we justify the non-principal-axis-rotation (NPAR) damping rate
given by Eq.~\eqref{eq:tau_npar}. While fully derived in \citet{peale1973}, the
origin of the result is somewhat obfuscated by the considerable scope of the
paper. Here, we provide two simple derivations that arrive very nearly at
Peale's result and provide a transparent physical origin of NPAR damping.

\subsection{Torque-Based Approach}

In this approach, the essential physical picture is to follow a standard
derivation of tidal obliquity damping \citep[we follow][]{2012Lai} but replacing
the tidal potential by the centrifugal potential induced by the planet's
rotation. For simplicity, we will specialize our discussion to an oblate body $B
= A$.

The figure of the planet experiences a centrifugal potential given by
\begin{equation}
    \Phi_{\rm cf} = -\frac{1}{2}\p{\bm{\Omega} \times \bm{r}}^2.
        \label{eq:app_phi_cf_0}
\end{equation}
Next, we express this potential in the spin frame (denoted with primes), where
the polar axis is parallel to the instantaneous angular velocity $\b{\Omega}$,
and we adopt the spherical coordinates $(r', \theta', \phi')$. In this
coordinate system, Eq.~\eqref{eq:app_phi_cf_0} becomes
\begin{align}
    \Phi_{\rm cf}(r', \theta', \phi') &= -\frac{\Omega^2(r')^2}{2}\sin^2\theta',\\
        &= \Omega^2(r')^2\frac{2\sqrt{\pi}}{3}
            \p{\frac{1}{\sqrt{5}}Y_{20}\p{\theta', \phi'} + Y_{00}}.
                \label{eq:app_phi_cf_s}
\end{align}
Here, we have introduced the $Y_{l'm'}\p{\theta', \phi'}$ spherical harmonics in
the spin frame.

The next step is to relate Eq.~\eqref{eq:app_phi_cf_s}, expressed in the spin
frame, to the torques experienced by the planet in its body frame (where the
polar axis is aligned with the principal axis of greatest moment of inertia).
The body and spin frames are misaligned by the angle $\beta$. This is typically
done by re-expressing Eq.~\eqref{eq:app_phi_cf_s} in terms of the body-frame
spherical harmonics $Y_{lm}\p{\theta, \phi}$ (unprimed coordinates denote the
body frame) and subsequently evaluating the torques by differentiating the
resulting potential. While nontrivial, this procedure is elementary and has been
thoroughly studied in the literature in calculating the equilibrium tide. We
will follow the notation and calculation of \citet{2012Lai} and simply adapt his
results. There, the $m' = 0$ component of the tidal potential generated by a
perturber of mass $M'$ at distance $a$ can be expressed in the coordinate system
where the perturber's orbit is in the equator ($\theta' = \pi/2$) as
\citep{2012Lai}\footnote{The sign of this term given in \citet{2012Lai} are
consistent with those given in \citet{tremaine_book}, but differ from some later
works. Note that the sign does not affect the physical torques, which only
depend on $W_{20}^2$.}:
\begin{equation}
    U_0(r', \theta', \phi') = GM'\sqrt{\frac{\pi}{5}}\frac{(r')^2}{a^3}
        Y_{20}\p{\theta', \phi'}.\label{eq:app_laiU0}
\end{equation}
\citet{2012Lai} then gives the components of the tidal torque in the
spin frame of the primary, which is misaligned by an angle $\Theta$ to the
orbital frame used in Eq.~\eqref{eq:app_laiU0}. By comparing
Eqs.~(\ref{eq:app_phi_cf_s},~\ref{eq:app_laiU0}), it is clear that his results
can be adapted to compute the components of the torque generated by our
spin-frame potential (Eq.~\ref{eq:app_phi_cf_s}) in the body frame (again,
which is misaligned from the spin frame by
$\beta$), by using the following
correspondence:
\begin{equation}
    \frac{GM'}{a^3} \Rightarrow \frac{2\Omega^2}{3}.\label{eq:app_correspondance}
\end{equation}
With this correspondence, we can then use the results of \citet{2012Lai} to
evaluate the resulting fictitious torque on the planet spin. Note that the
fictitious torque contains a contribution oriented along $\khat \times
\bm{\Omega}$ that would contribute to tidally-induced spin precession
and thus must vanish at leading order. We choose our coordinate system such that
$\jhat \parallel \khat \times \bm{\Omega}$. The remaining components of the
torque along the $\khat$ and $\ihat$ body axes are:
\begin{align}
    T_k &= \frac{3\pi}{5}T_0\Omega\p{\sin^4\beta\tau_{20}
            + \sin^2\beta \cos^2\beta\tau_{10}},\label{eq:T_k}\\
    T_i &= -\frac{3\pi}{5}T_0\Omega\p{
        \sin \beta^3\cos\beta\tau_{20}
        + \sin\beta\cos^3\beta \tau_{10}},\label{eq:T_i}
\end{align}
where
\begin{equation}
    T_0 = \frac{4\Omega^4}{9}\frac{R^5}{G}.
\end{equation}
Here, $\tau_{mm'}$ refers to the tidal lag time corresponding to the $(mm')$
component of the tidal potential, and $\beta$ appears as the angle between the
two frames. Note that these have the opposite sign to the expressions given in
Eqs.~(27, 35) of \citet{2012Lai}, which are the torques exerted \emph{by} the
perturbing potential, while we need the torque exerted \emph{on} the perturbing
potential (generated by the planet's spin angular momentum). This sign choice
ensures that NPAR damps, rather than grows. Finally, note that these components
are exactly the components of $\bm{\Gamma}_{\rm NPAR}$ as given in
Eq.~\eqref{eq:npar_torque}.

Next, we need to relate the torques to the evolution of the NPAR angle $\cos
\beta = {\khat} \cdot \uv{\Omega} = \Omega_k/\Omega$. For oblate bodies,
we can explicitly evaluate (using the spin evolution
Eqs.~\ref{eq:euler1}--\ref{eq:dijk-dt})
\begin{align}
    \rd{}{t}\cos \beta
        &= \frac{T_k(\Omega^2 - \Omega_k^2)}{C\Omega^3}
        - \frac{\Omega_k\Omega_iT_i}{A \Omega^3}.\label{eq:dbeta}
\end{align}

Substituting Eqs.~(\ref{eq:T_k}--\ref{eq:T_i}) into Eq.~\eqref{eq:dbeta}, we
obtain
\begin{align}
    \rd{}{t}\cos\beta
        ={}& \frac{4\pi \Omega^4R^5}{15GC}\s{
            (\tau_{20}\sin^6\beta + \tau_{10}\sin^4\beta\cos^2\beta)
            + \frac{C}{A}\p{\tau_{20}\sin^4\beta\cos^2\beta
                + \tau_{10}\sin^2\beta\cos^4\beta}
            }.\label{eq:app_full_dbetadt}
\end{align}
For small angles $\beta$, the dominant term is
\begin{align}
    \rd{\beta}{t} \approx{}&
        -\frac{4\pi \Omega_k^3\Omega}{15}\tau_{10}
            \frac{R^5}{GA}\sin\beta.
\end{align}
Defining that $k_2/Q = 4\pi \Omega \tau_{10} / 5$ \citep{2012Lai}, we write
\begin{align}
    \rd{\beta}{t} &\approx -S\frac{k_2}{Q}\Omega_k^3
            \frac{R^5}{3GA}\sin\beta \label{eq:app_npar_ysfirst}\\
        &\approx -\frac{1}{3k}\frac{k_2}{Q}\Omega_k^3
            \frac{R^3}{GM}\sin\beta.
\end{align}
In the second line, we've taken $k \equiv C/MR^2 \approx A/MR^2$.
The first expression (Eq.~\ref{eq:app_npar_ysfirst}) differs from the expression
in \citet{peale1973} by a factor of $3A / (2A + C)$ (both results are for oblate
bodies only).

Thus, we conclude that the simple physical picture where a deformation of the
planet body that tracks the centrifugal potential in a time-lagged manner is
sufficient to closely reproduce the classical expression from \citet{peale1973}.
As in \citet{2012Lai}, we note that our calculation can accommodate more
realistic tidal models by parameterizing the dissipation in terms of a tidal
time lag (or equivalently, $Q$) that depends on $\Omega$ and the various other
properties of the system \citep[e.g.\ that of][]{Yoder1979, efroimsky2001}.

\subsection{Energy-Based Approach}

In this section, we will obtain the NPAR damping rate entirely from first
principles via a careful analysis of the energetics of the system. For
simplicity we assume
\begin{enumerate}
  \item An oblate or nearly oblate body: $0\le B-A\ll C-A$;
   \item A small amplitude for the non-PAR, i.e. $\Omega_k\gg\Omega_i,\Omega_j$;
   \item Free precession, i.e.\ we ignore external torques due to tides etc.
   \end{enumerate}
Under these assumptions, the Euler equations for free rotation expressed in the
body frame
\begin{align}
 I_a\dot\Omega_a&=\sum_{b,c\,\in\{i,j,k\}} \tfrac{1}{2}\epsilon_{abc}(I_b-I_c)\Omega_b\Omega_c\,,\nonumber\\
 \mbox{imply that}\qquad\qquad
 \Omega_\perp(t)&\equiv \Omega_i(t)+i\Omega_j(t) = \Omega_\perp(0) e^{i\Wnpar t}\,,
\end{align}
in which $\Omega_\perp$ is the amplitude of the NPAR---treated as a
complex variable---and
\begin{equation}
  \label{eq:Omega}
  \Wnpar = \Omega_k\sqrt{\frac{(C-A)(C-B)}{AB}}\approx \frac{C-A}{A}\,\Omega_k
\end{equation}
is its frequency, i.e.\ the wobble frequency.
NPAR gives rise to time-dependent elastic stresses in the body, which
damp at a rate $\propto Q^{-1}$.
The timescale Eq.~\eqref{eq:tau_npar} can be understood as
$(d\ln|\Omega_\perp|/dt)^{-1}$.

The oscillatory part of the
centrifugal potential due to NPAR (Eq.~\ref{eq:app_phi_cf_0}) is quadrupolar:
\begin{align}\label{eq:poten}
  -\tfrac{1}{2}(\b{\Omega\times r})^2 &= \Omega_i\Omega_j r_ir_j
  -\tfrac{1}{2}(\Omega_k^2+\Omega_j^2)r_i^2 +[\mbox{cyclic permutations
                                        $(ijk)\to(jki)$}]\nonumber\\[1ex]
  &\approx \Omega_k r_k(\Omega_i r_i + \Omega_j r_j)\nonumber\\
    &= \mbox{Real}\left[-\sqrt{\frac{8\pi}{15}} \Omega_k \Omega_\perp(0) \, r^2
                                                 Y_{21}(\theta,\,\phi-\Wnpar t)\right]\,;
\end{align}
in the second line, only the terms 1st-order in
$\Omega_\perp$ have been retained per assumption 2.

The distortion due to the wobble-induced stresses have the same form as for a
true quadrupolar tide due to an external mass $M_\mathrm{ext}$ at some distance
$D\gg R$. The instantaneous potential induced by such a mass is given by
\begin{equation}\label{eq:P2}
  \delta\Phi_\mathrm{ext}(r,\theta) = -\frac{GM_\mathrm{ext}}{D^3} r^2 P_2(\cos\theta)
= -\sqrt{\frac{4\pi}{5}}\,\frac{GM_\mathrm{ext}}{D^3} r^2Y_{20}(\theta,\phi)
\end{equation}
in polar coordinates centered on the elastic body with $\theta=0$ in the
direction toward $M_{\rm ext}$. Note that this differs from
Eq.~\eqref{eq:app_laiU0} because the perturber is placed at $\theta=0$ rather
than at $\theta = \pi / 2$. If the undistorted elastic body were
spherical,\footnote{But of course it is not, since $C>B\ge A$, so
Eq.~\eqref{eq:Estored} is only approximate.} its responding distortion would
also be proportional to $P_2(\cos\theta)$, so that
$\delta\Phi_\mathrm{resp}(D,0)=k_2\delta\Phi_\mathrm{ext}(R,0)(R/D)^3$. By
considering the work done on $M_\mathrm{ext}$ by $\delta\Phi_\mathrm{ext}$ as
$D$ is brought down from infinity, one sees that the elastic-plus-potential
energy in the distortion is $\delta E_\mathrm{self} =
k_2GM_\mathrm{ext}^2R^5/2D^6$. Replacing the coefficient of $Y_{2,0}$ in
Eq.~\eqref{eq:P2} with that of $Y_{2,1}$ in eq.~\eqref{eq:P2} yields
\begin{equation}\label{eq:Estored} \delta E_\mathrm{self} \approx
k_2\frac{\Omega_k^2|\Omega_\perp|^2 R^5}{3G} \end{equation} for the self-energy
of the (time-dependent part of) the centrifugal distortion.

By definition of $Q$, the dissipation rate averaged over a cycle (period
$2\pi/\Wnpar$) is\footnote{$Q$ is defined in terms of the \emph{peak} stored
potential energy, which would be twice the mean stored energy for a harmonic
oscillation in a single degree of freedom.  However,
$|\omega_\perp|^2=\Omega_i^2+\Omega_j^2$ is approximately constant because
$\Omega_i$ and $\Omega_j$ are $90^\circ$ out of phase and have similar
amplitudes, so the energy (Eq.~\ref{eq:Estored}) doesn't vary much (or at all,
if $A=B$) over a cycle.}
\begin{equation}
  \label{eq:Edisp}
  \delta\dot E_\mathrm{self} = \frac{\Wnpar}{Q} \delta E_\mathrm{self}
\end{equation}
However, in our case, the damping time is not simply $\delta
E_\mathrm{self}/\delta \dot E_\mathrm{self}$, because the relevant stored energy
is not the elastic-plus-potential self-energy $\delta E_{\rm  self}$, but rather
the \emph{kinetic} energy of the NPAR\@. The latter is the difference between
the actual rotational kinetic energy and the kinetic energy of principal-axis
rotation at the same angular momentum. The square of the spin angular momentum
is (for $A\approx B<C$)
\begin{equation*}
  S^2 \approx  A^2|\Omega_\perp|^2 + C^2\Omega_k^2\,,
\end{equation*}
and the total spin kinetic energy is
\begin{equation*}
  E_\mathrm{spin} \approx  \frac{1}{2}\left(A|\Omega_\perp|^2 + C\Omega_k^2\right)\,,
\end{equation*}
and the energy if $\b{S}$ were aligned with the principal axis would
be $S^2/2C$, so the kinetic energy in the wobble is
\begin{align}
  \label{eq:Enpar}
  E_{\textsc{npar}} &\approx \frac{1}{2}\left(A|\Omega_\perp|^2 + C\Omega_k^2\right)-
\frac{1}{2C}\left(A^2|\Omega_\perp|^2 + C^2\Omega_k^2\right)\nonumber\\[1ex]
&= \frac{A(C-A)}{2C}|\Omega_\perp|^2
\end{align}
Therefore, by combining Eqs.~(\ref{eq:Omega},~\ref{eq:Estored},~\ref{eq:Edisp},
and~\ref{eq:Enpar}),
the damping time of the NPAR amplitude (twice that of $E_{\textsc{npar}}$) is seen to be
\begin{align}
  \label{eq:tdamp}
  \tnpar &= 2\frac{E_{\textsc{npar}}}{\delta\dot E_\mathrm{self}} =
            2\frac{Q E_{\textsc{npar}}}{\Wnpar\delta E_\mathrm{self}}
\approx \frac{3Q}{k_2}\frac{GA^2}{CR^5}\Omega_k^{-3}\,,
\end{align}
Since $\Omega_k=\dot\psi$ (the component of angular velocity along the principal
axis), this differs from the expression given by \citet{peale1973} only by a
factor $3A^2/C(2A+C)$, which would be unity for a spherical body; on the other
hand, the use of $k_2$ is strictly justified only for a sphere.

\bibliographystyle{aasjournal}
\bibliography{refs}{}

\end{document}